\documentclass[sigconf, screen]{acmart}
\usepackage{amsmath}
\usepackage{hyperref}
\usepackage{cleveref}
\usepackage{algorithmic}
\usepackage{graphicx}
\usepackage{textcomp}
\usepackage{xcolor}
\usepackage{float}
\usepackage{multirow}
\usepackage{booktabs} 
\usepackage{adjustbox}
\usepackage{tcolorbox}
\usepackage{balance}

\newcommand{\CompanyX}{Microsoft}
\renewcommand\footnotetextcopyrightpermission[1]{}
\acmBooktitle{Companion Proceedings of the 33rd ACM Symposium on the Foundations of Software Engineering (FSE '25), June 23--27, 2025, Trondheim, Norway}
\AtBeginDocument{%
  }

\begin{document}

\title{eARCO: Efficient Automated Root Cause Analysis with Prompt Optimization}

\author{
   Drishti Goel, Raghav Magazine, Supriyo Ghosh, Akshay Nambi, Prathamesh Deshpande, Xuchao Zhang, Chetan Bansal, Saravan Rajmohan \\
    {\large Microsoft}
}

\begin{abstract}
Root cause analysis (RCA) for incidents in large-scale cloud systems is a complex, knowledge-intensive task that often requires significant manual effort from on-call engineers (OCEs).  Improving RCA is vital for accelerating the incident resolution process and reducing service downtime and manual efforts. Recent advancements in Large-Language Models (LLMs) have proven to be effective in solving different stages of the incident management lifecycle including RCA. However, existing LLM-based RCA recommendations typically leverage default finetuning or retrieval augmented generation (RAG) methods with static, manually designed prompts, which lead to sub-optimal recommendations. 
In this work, we leverage `PromptWizard', a state-of-the-art prompt optimization technique, to automatically identify the best optimized prompt instruction that is combined with semantically similar historical examples for querying underlying LLMs during inference.  
Moreover, by utilizing more than 180K historical incident data from \CompanyX{}, we developed cost-effective finetuned small language models (SLMs) for RCA recommendation generation and demonstrate the power of prompt optimization on such domain-adapted models.  
Our extensive experimental results show that prompt optimization can improve the accuracy of RCA recommendations by 21\% and 13\% on 3K test incidents over RAG-based LLMs and finetuned SLMs, respectively. Lastly, our human evaluation with incident owners have demonstrated the efficacy of prompt optimization on RCA recommendation tasks.
These findings underscore the advantages of incorporating prompt optimization into AI for Operations (AIOps) systems, delivering substantial gains without increasing computational overhead.
\end{abstract}


\maketitle
\pagestyle{plain}

\section{Introduction}
Over the past decade, large-scale cloud services have become essential for deploying and maintaining complex interdependent systems. Despite significant reliability efforts, these services still experience production incidents like unplanned outages or performance degradations, leading to customer dissatisfaction, revenue loss, and declining trust. The current incident diagnosis process heavily relies on manual efforts from on-call engineers (OCEs), resulting in productivity loss. Additionally, limited expertise or domain knowledge among OCEs can lead to sub-optimal actions, delaying service recovery.

Root cause analysis (RCA) is a critical and time-consuming step in the incident management lifecycle, requiring deep domain knowledge and back-and-forth communication among OCEs. OCEs refer to several information source such as troubleshooting guides, past similar incidents, service properties and current incident metadata to identify the key reasons behind service disruptions. Accurately identifying the root cause early can significantly reduce time-to-mitigate (TTM) by enabling faster execution of mitigation steps. Thus, automating RCA at an early stage could accelerate incident resolution, reduce service downtime, and minimize manual efforts by OCEs.

Recent advancements in LLMs have proven to be effective for solving several problems in the incident management lifecycle, ranging from detection \cite{ganatra2023detection,Srinivas2024IntelligentMonitoring} to triaging \cite{azad2022picking,chen2019continuous} to automated query recommendations~\cite{jiang2023xpert} to problem categorization \cite{chen2023empowering} to root cause generation \cite{ahmed2023recommending,zhang2024automated,goel2024x}. For root cause generation, \cite{ahmed2023recommending} first propose to finetune a GPT-3 model with initial incident metadata (e.g., title, initial summary and owning service name) as input and the corresponding root cause as output in a supervised setting. However, finetuning LLMs is computationally expensive and maintenance-heavy.  \cite{zhang2024automated} proposed a retrieval-augmented generation (RAG) based in-context learning (ICL) solution that dynamically retrieves similar historical incidents during inference and prompts a pre-trained LLM with predefined, manually designed instructions. While this RAG-based ICL method has demonstrated potential, significant challenges arise when scaling it for RCA: \textbf{(1) Static Prompts:} Manually defined prompts lack the flexibility to adapt as tasks and models evolve, making continual updates labor-intensive. \textbf{(2) Sub-optimal Guidance: }Manually crafted prompts may not fully leverage the LLM’s potential, leading to sub-optimal RCA recommendations without automated optimization. \textbf{(3) Cost and Scalability:} While LLMs like GPT-4 are powerful, they are costly to deploy at scale. This raises the need for more cost-effective solutions, such as fine-tuning smaller language models (SLMs), without compromising on RCA accuracy.

In this paper, we introduce a novel framework for \textbf{e}fficient \textbf{a}utomated \textbf{r}oot \textbf{c}ause analysis with prompt \textbf{o}ptimization (\textbf{eARCO}). Our work addresses the following key research questions: (1) \textbf{RQ1}: Can automatically optimized prompt instructions outperform manually designed static instructions in improving the accuracy and quality of RCA recommendations? (2) \textbf{RQ2}: Does the combination of optimized prompt instructions and strategically selected in-context examples deliver superior RCA performance compared to using prompts or examples alone? and (3) \textbf{RQ3}: Can SLMs, when paired with optimized prompt instructions, provide a cost-effective alternative to querying expensive LLMs while maintaining comparable RCA performance?

To address these research questions, we leverage state-of-the-art prompt optimization techniques to automatically generate optimized prompt instructions tailored for the RCA task. The process begins by taking a task description and a few training examples of incidents with their root causes, after which optimized prompts are derived. We then select in-context learning (ICL) examples that closely match the current incident.

Specifically, we leverage PromptWizard \cite{agarwal2024promptwizard}, a discrete prompt optimization approach that evolves and adapts its prompts in real-time. PromptWizard has shown superior performance across various NLP tasks by generating, critiquing, and refining prompts through an iterative feedback loop. This "critic-and-synthesize" process continuously improves prompts without requiring additional model training. By querying a pre-trained LLM fewer than 100 times, the method remains computationally efficient while producing highly optimized prompt instructions for RCA.
This automated prompt optimization eliminates the need for manual prompt engineering, ensuring that as pre-trained LLMs evolve, the     system consistently provides optimal instructions for improving model performance. 
Once the optimized instruction is obtained, we develop a retrieval-augmented generation (RAG) system inspired by Zhang et al. \cite{zhang2024automated} to dynamically retrieve the top-K semantically similar historical incidents and their root causes. 
By incorporating these past incidents and their root causes, the system adds valuable domain-specific context, enhancing the model's ability to generate accurate RCA recommendations.
Our extensive experimental results, tested on 2,900 real-world incidents from \CompanyX{}, demonstrate that combining optimized prompt instructions with semantically similar in-context examples significantly improves the quality and accuracy of RCA recommendations.

Despite the strong performance of large language models (LLMs) for RCA tasks, their use in production with long context lengths is prohibitively expensive. To address this, we fine-tune smaller models like Phi-3-Mini, Phi-3-Medium \cite{abdin2024phi}, and Phi-3.5-Mini using 180K historical incidents from \CompanyX{}, incorporating metadata and root causes from over 1,000 services. Combining these fine-tuned SLMs with optimized prompts from PromptWizard improves RCA accuracy by 13\% compared to using incident context alone. This demonstrates that fine-tuned SLMs with optimized prompts provide a cost-effective alternative to LLMs, reducing inference costs and complexity.

This paper makes three key contributions:
\begin{itemize}
    \item We introduce \textbf{eARCO}, a novel framework that integrates optimized prompts from PromptWizard with top-K semantically similar historical incidents to generate accurate RCA recommendations, dynamically adapting to task-specific needs. 
    \item We develop a cost-effective RCA solution by fine-tuning SLMs on 180K historical incidents from over 1,000 services at \CompanyX{}. These SLMs, when queried with optimized prompts, offer a cost-effective alternative to expensive LLMs without compromising performance. 
    \item GPT-4-based evaluations show that prompt optimization improves RCA accuracy by 21\% for LLMs and 13\% for SLMs. Human evaluations with domain experts confirm the practical benefits, reporting enhanced RCA task performance.
\end{itemize}

\section{Background}
In this section, we begin by introducing the incident management domain and providing background on root cause analysis (RCA). Next, we discuss the potential advancements in incident management enabled by recent developments in large language models (LLMs). This is followed by outlining our key research questions. Lastly, we detail the data preparation strategy for the RCA task, which includes steps for data curation, cleaning, and summarization. 

\subsection{Incident Root Cause Analysis}
Despite significant reliability efforts, large-scale cloud services inevitably encounter production incidents or outages, which can result in significant customer impact and financial loss. On-call engineers (OCEs) need extensive domain knowledge and expend considerable manual effort to diagnose and resolve these incidents. The incident lifecycle typically involves four stages:
\begin{itemize}
    \item \textbf{Detection:} Incidents are detected either by external/internal service users or through automated monitoring systems set up to track service health and performance. 
    \item \textbf{Triaging:} Once detected, incidents are routed to the appropriate service teams or OCEs based on the incident properties and team expertise.
    \item \textbf{Root cause analysis:} OCEs engage in rounds of communication, analyzing logs, performance metrics, service dependencies, and troubleshooting guides to identify the root cause. 
    \item \textbf{Mitigation:} Mitigation actions are taken based on the identified root cause to resolve the incident and restore service functionality.
\end{itemize}
Root cause analysis is particularly challenging, requiring significant manual effort and domain knowledge. Incidents may arise from various sources, such as code bugs, configuration errors, dependency failures, or hardware issues. Missteps in RCA can delay service recovery and exacerbate customer impact. Automating the identification of potential root causes early in the incident lifecycle can guide OCEs toward the correct resolution path, reducing overall time-to-mitigate (TTM) and minimizing customer disruption. 

\subsection{Promise of LLMs in Incident Management}
The rapid advancements in LLMs have led to exceptional performance in a wide variety of natural language tasks, ranging from summarization and translation to question-answering and code completion. Furthermore, recent advancements in domain adaptation using finetuning and few-shot learning have laid the foundation for solving problems in different stages of the incident lifecycle. LLMs have shown great potential in automating detection, problem categorization and accurate triaging of incidents which can alleviate the load and manual efforts from engineers. Moreover, recent studies \cite{ahmed2023recommending, zhang2024automated, goel2024x} have demonstrated the usefulness of LLMs in automatically identifying the root causes of an incident by leveraging initial incident metadata and similar historical incident properties, either by finetuning LLMs or using in-context (ICL) learning framework. However, the use of static, manually designed prompt instructions can result in sub-optimal performance, and running LLMs with long context lengths for ICL remains costly. To address these limitations, we propose using automatically optimized prompt instructions, coupled with cost-effective fine-tuned smaller language models (SLMs), to enhance root-cause generation accuracy while minimizing operational costs.

\subsection{Research Questions} Our goal is to address the following three main research questions in this study:
\begin{itemize}
\item \textbf{Can optimized prompt instruction improve the quality of RCA recommendation?}: Manually designing high-quality static prompts is challenging and often leads to sub-optimal results. This requires significant domain knowledge and is mostly trial-and-error. Additionally, performance may degrade as LLM models or their versions evolve. We aim to leverage the state-of-the-art prompt optimization technique (`PromptWizard') to automatically identify the best prompt instructions and demonstrate their superiority over manually crafted prompts. 
 \item \textbf{What role do in-context examples play in conjunction with optimized instructions?} The effectiveness of LLMs in reasoning tasks largely depends on the quality of in-context examples included in the prompt. We will evaluate the performance of static in-context examples identified by `PromptWizard' against dynamically retrieved semantically similar examples. 
 \item  \textbf{How do cost-effective finetuned SLMs perform with optimized prompts?} To reduce reliance on expensive LLMs with large context lengths, we fine-tune SLMs using historical incident data. We will assess the performance of optimized instructions generated by `PromptWizard' on these fine-tuned SLMs.  
\end{itemize}

\subsection{Data Preparation} To address the research questions on the RCA task, we curated a comprehensive incident dataset from \CompanyX{}. The dataset includes over 180K historical incidents and their corresponding root causes. We performed data cleaning and pre-processing to ensure consistency, removed any incomplete or irrelevant records, and summarized the key incident properties.

\subsubsection{Data Collection}
Each service team at \CompanyX{} logs their incident data in the internal incident management (IcM) portal. We curated a dataset of approximately 180K high-severity historical incidents from January 2022 to June 2024 from this portal. For each incident, we collected key metadata, including the title, initial summary (written at the time of the incident), owning service name, and the ground truth root cause identified and documented by the on-call engineers (OCEs). This data serves as the basis for both fine-tuning models and evaluating root cause analysis performance.

\subsubsection{Data Cleaning and Summarization}
The raw summaries and root causes of incidents collected from the IcM portal often contain significant amounts of extraneous information, such as HTML tags, images, stack traces, and code snippets. This noise can adversely impact model performance during both training and testing for several reasons: (a) the effectiveness of a fine-tuned model relies heavily on high-quality training data; noise diminishes the model's ability to learn effectively; (b) excessive noise can impair the reasoning capabilities of both base and finetuned models, potentially leading to hallucinations; and (c) accurate ground truth for root causes is essential for evaluating model effectiveness.

To mitigate such interference, we employed a two-stage data cleaning strategy. First, we locally processed the data to remove irrelevant HTML tags, stack traces, and image tags. In the second stage, we utilized the GPT-3.5-turbo\footnote{\href{https://platform.openai.com/docs/models/gpt-3-5-turbo}{https://platform.openai.com/docs/models/gpt-3-5-turbo}} model to summarize the root cause and summary fields, guided by a prompt as proposed by Zhang et al.~\cite{zhang2024automated}. This summarization step also enabled the LLM to identify and filter out noisy or non-informative root causes, enhancing the overall quality and stability of the dataset.

\section{Prompt Optimization for RCA} \label{sec:promptwizard}

We now provide the details of our \textbf{eARCO} framework that primarily consists of two components: \textit{Prompt instruction optimization} and \textit{In-Context example selection}. We begin by explaining each of these two components while emphasizing their respective roles in optimizing the quality of RCA recommendations. We then explain the overall architecture of the eARCO system by integrating these two components together. 


\begin{figure*}[!htb]
\begin{tcolorbox}[width=\linewidth, title={Optimized prompt instruction for RCA.}]
You will be given a detailed description of an incident involving a system or service, including pertinent logs, error messages, and other relevant information. Your objective is to methodically analyze the incident and identify the root cause. Follow these steps:\\
1. \textbf{**Contextual Information**:} Identify the service, relevant timestamps, environment, and key stakeholders involved. Mention regional specifics where applicable.\\
2. \textbf{**Categorization**:} Categorize the type of incident (e.g., compute issues, storage conflicts, network anomalies).\\
3. \textbf{**Identify Symptoms**:} List all symptoms and error messages mentioned in the incident description.\\
4. \textbf{**Detailed Historical Review**:} \\
    - Reflect on similar incidents and any historical data that might provide insights.\\
    - Explicitly assess any past configuration changes or known historical script configurations related to the issue.\\
5. \textbf{**Environmental Variables and Changes**:}\\
    - Identify and evaluate recent environmental changes, including recent configuration updates or external factors.\\
    - Refer to specific timestamps leading up to and during the incident for environmental and systemic changes.\\
6. \textbf{**Analyze Patterns and Logs**:}\\
    - Examine logs and error messages for recurring patterns.\\
    - Cross-verify error logs against configuration settings and recent system changes.\\
    - Look for specific script logs or monitor configurations and validate against system norms.\\
7. \textbf{**Root Cause Analysis**:} \\
    - Synthesize findings from logs, historical data, and environmental variables.\\
    - Clearly delineate between potential and confirmed root causes.\\
    - Loop back to compare symptoms with broader historical and configuration data to ensure comprehensive scrutiny.\\
8. \textbf{**Conclusion**:} Clearly present the final root cause(s) wrapped between $<ANS\_START>$ and $<ANS\_END>$ tags.\\
Be thorough and evidence-based in your analysis, while eliminating any personal biases. Base your findings entirely on the provided details to ensure accuracy.
\end{tcolorbox}
\caption{Optimized prompt instruction identified by PromptWizard.}
\label{fig:optimizedprompt}
\end{figure*}
\subsection{Prompt Optimization}
LLM responses are highly sensitive to prompt instructions, making it crucial to provide the right instructions for generating high-quality RCA recommendations. To achieve this, we leverage a state-of-the-art prompt optimization technique called \textit{PromptWizard} (PW) \cite{agarwal2024promptwizard}. PW is a discrete optimization approach that refines manually designed initial prompts using input-output pairs from the training data. The optimization process involves four key steps: Mutate, Score, Critique, and Synthesize. Each step employs LLMs to execute pre-defined actions through specialized prompt templates. These steps are run iteratively to progressively optimize both the prompt instructions and the selection of in-context examples.

\textbf{Prompt Instruction Tuning :}
In the \textbf{Mutate} stage, Prompt Wizard (PW) takes the initial prompt instructions and task description, applying predefined thinking styles to generate prompt variations in a single LLM call. In the \textbf{Score} stage, these mutated prompts are evaluated on a diverse batch of training samples, with each prompt assigned a score based on performance. The best-performing prompt proceeds to the \textbf{Critique} stage, where it receives targeted feedback regarding its strengths and weaknesses. This feedback is used in the \textbf{Synthesize} stage to further refine and improve the prompt. The refined prompt is then fed back into the Mutate stage, continuing iteratively until either a predefined performance threshold is met or the maximum number of iterations is reached. This feedback-driven loop effectively balances exploration and exploitation, continuously enhancing the prompt.\\
\textbf{In-Context Examples Tuning :}
The inclusion of in-context examples (ICL) is optional for users. If selected, the optimized prompt from the previous phase is further tuned along with ICL examples. PW begins by selecting a random set of 25 diverse examples, covering both positive and negative instances. These examples are optimized in two phases: First, the \textbf{Critique} stage provides feedback on how well the examples complement the prompt instructions, suggesting improvements. Then, in the \textbf{Synthesize} stage, this feedback is used to refine the instructions and example selection. Second, PW identifies the optimal mix of positive, negative, and synthesized examples through this iterative critique-synthesis loop, adapting both instructions and examples to the task at hand.

To further enhance performance, the prompt enters a \textbf{Reasoning} stage, where chain-of-thought (CoT) reasoning is incorporated into the examples, followed by a \textbf{Validation} stage to verify the correctness of the examples and reasoning, preventing hallucination and removing errors. PW also introduces a \textit{Task Intent} and an \textit{Expert Persona} to the optimized prompt. The Task Intent helps the LLM maintain relevance, while the Expert Persona ensures consistency with domain expertise.

The final optimized prompt consists of four main components: a problem description, optimized instructions, static and diverse in-context examples, and task intent with an expert persona. Optionally, an \textit{answering format} section can be added to specify how the LLM should structure its response for downstream tasks. Importantly, this prompt optimization is a one-time process; the optimized prompt is then reused at the inference stage for all incident analyses.
\\
\textbf{Adapting to RCA Task :}
For our RCA task, we begin by selecting 25 to 30 diverse historical incident examples from our training corpus. These examples are chosen to ensure coverage across various root-cause categories. We choose a manually designed prompt instruction from Zhang \emph{et. al.} \cite{zhang2024automated}, as the initial input to PromptWizard. PW then generates an optimized prompt instruction for RCA task and identifies a set of diverse in-context examples. The \textit{Expert Persona} guides the model to take on the role of an On-Call Engineer (OCE) with advanced analytical and reasoning skills, tasked with identifying the root cause of cloud system incidents. To improve the performance of PW optimization modules, we tune the following hyper-parameters and its values are described in Section~\ref{sec:conf}: 
\begin{itemize}
    \item \textit{mutate\_refine\_iterations} : Number of iterations for conducting rounds of mutation of task description and refinement of instructions.
    \item \textit{mutation\_rounds} : Number of rounds of mutation to be performed when generating different styles.
    \item \textit{refine\_task\_eg\_iterations} : Number of iterations for refining task description and in-context examples.
    \item \textit{questions\_batch\_size} : Number of questions to be asked to LLM in a single batch during training.
    \item \textit{min\_correct\_count} : Number of batches of questions to be correctly answered, for a prompt to be considered for the next steps.
    \item \textit{few\_shot\_count} : Number of in-context examples in the final prompt.
\end{itemize}
These parameters govern the extent of optimization and the final structure of the prompt. The optimized prompt instructions identified by PromptWizard for the RCA task, using the GPT-4o model, are illustrated in Figure~\ref{fig:optimizedprompt}. A crucial aspect of PW's optimization process is the iterative refinement of both the instruction and the in-context examples. During this stage, knowledge from the examples informs the refinement of the instructions, and vice versa. An example of this in Figure~\ref{fig:optimizedprompt} includes the line: \textit{Categorization: Categorize the type of incident (e.g., compute issues, storage conflicts, network anomalies)}. Here, the examples list is generated based on information derived from the training data, highlighting the knowledge transfer between the two prompt components. This back-and-forth refinement leads to the optimal setting for both the instructions and in-context examples. Moreover, the generated prompt is structured in a clear, step-by-step format, guiding the model from the incident information through critical analysis and ultimately to the root cause. By leveraging the training data, PromptWizard produces a task-specific prompt that is highly effective for RCA, ensuring alignment with the problem domain.

\subsection{In-context example selection}
\label{subsection:ICL}


In-context examples play a vital role during LLM inference, especially for RCA task as the pre-trained LLMs do not have complete knowledge of the incident management domain during training.  
Although PW generates a set of static in-context examples which shown superior performance on traditional NLP tasks, the context for each incident typically varies significantly. Therefore, a set of static examples may not be suitable for all incidents.  

To dynamically understand the current incident context and identify semantically similar historical examples, we used a traditional retrieval augmented generation (RAG) pipeline. For each of the 180K incidents in training corpus, we first combine the title and cleaned summarized version of initial incident summary, and encode them into vector representations using the Sentence Transformer model \cite{karpukhin-etal-2020-dense}. All these training incidents' metadata along with the corresponding embedding value are then stored in a vector database. 


Once the embedding vectors are indexed, we leverage the FAISS library \cite{johnson2019billion} for the efficient similarity search and clustering of the dense vectors. 
The FAISS library utilizes a compressed representation of the vectors, which eliminates the need to store the original vectors in memory. While this compression may result in a marginal reduction in search precision, the key advantage lies in its exceptional scalability. FAISS can efficiently handle billions of vectors in main memory on a single server. For a given incident and its text embedding vector during inference, FAISS then retrieves the top-K similar incidents using the L2 (Euclidean) distance metrics.

\begin{figure*}[ht]
    \centering
    \includegraphics[width=0.75\linewidth]{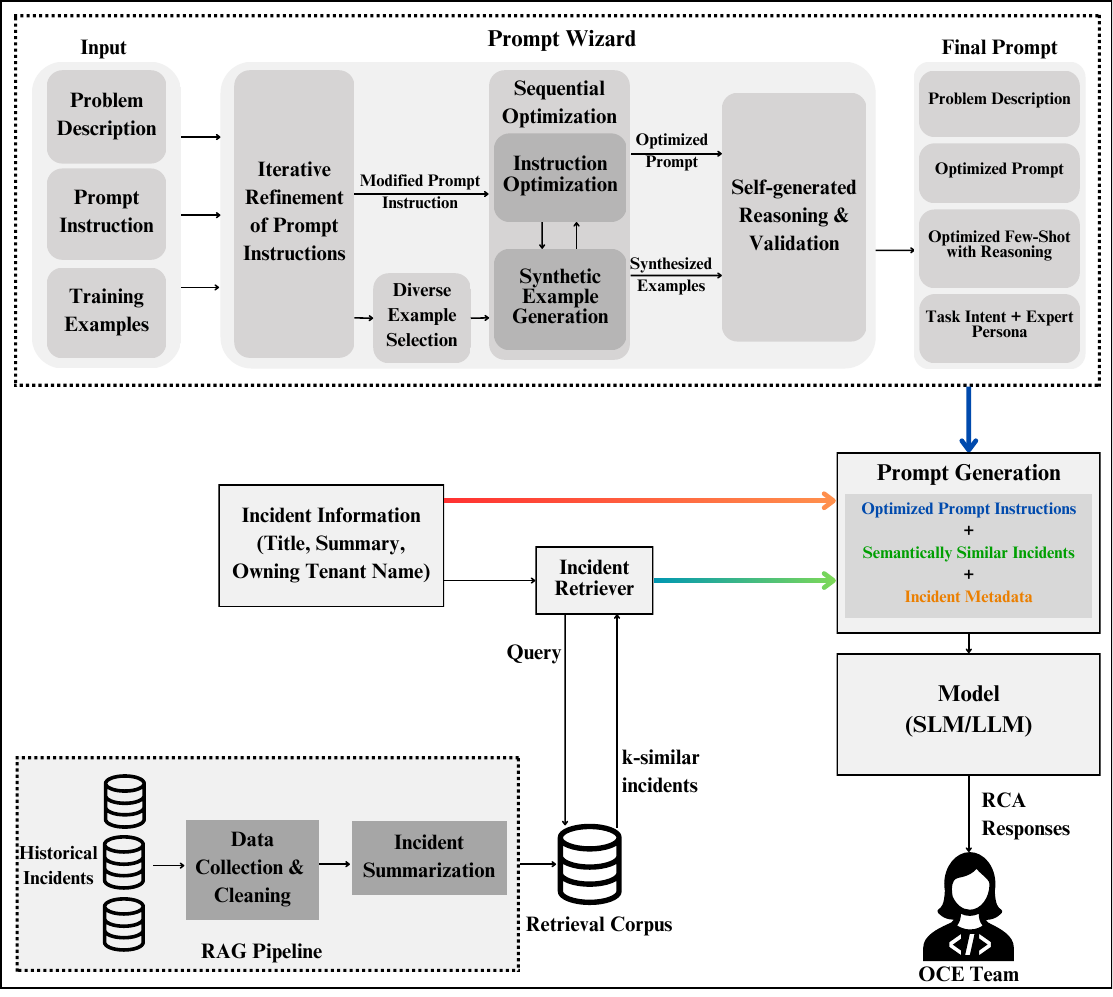}
    \caption{Architecture of the eARCO Framework for Efficient Root Cause Analysis (RCA)}
    \label{fig:Arch}
\end{figure*}

\subsection{Leveraging eARCO for RCA Generation}

In this paper, we propose \textbf{eARCO}, a comprehensive framework for efficient automated root cause analysis (RCA). eARCO combines the \textit{Prompt Instruction Optimization} and \textit{In-context example selection} techniques to significantly enhance the performance of LLMs in generating accurate RCA recommendations. The detailed architecture of this framework is illustrated in \cref{fig:Arch}. 

There are two stages of eARCO, (1) Optimized prompt instruction generation one-time and (2) Selection of ICL examples at inference time. Specifically, at inference time, given the current incident metadata (i.e., title and initial summary which are available at the time of incident creation), we generate the encoded embedding vector using the Sentence Transformer model which is used as an incident query vector. Using the retrieval pipeline outlined in \cref{subsection:ICL}, eARCO then dynamically selects the top 10 \textit{semantically similar incidents} from the historical incident corpus. 

The semantic similar examples are augmented with the static optimized prompt instruction computed using PW (as outlined in \cref{sec:promptwizard}). 
By integrating the \textit{optimized instructions}, \textit{incident metadata}, and \textit{semantically similar examples}, eARCO enhances the capabilities of LLMs to produce better quality automated RCA recommendations without additional training. 

\section{eARCO for Finetuned SLMs}
Another significant contribution of this paper is our exploration of finetuned small language models, optimized eARCO framework.
In contrast to relying on retrieval-based methods to supply language models with historical incidents for domain-specific tasks like root cause analysis, fine-tuning offers a more targeted and efficient approach. Finetuning eliminates the need for maintaining a large retrieval corpus, large contextual information in prompts as ICL examples, inference time computations and significantly reduces the risk of hallucinations. Previous efforts, of finetuning the GPT-3 model, have demonstrated that adapting language models for industry-specific tasks can be highly effective, but as previously mentioned, finetuning and maintaining large language models is a costly task. Our work addresses this by employing small language models (SLMs) of varying sizes, which not only optimize performance for specialized tasks but also presents a resource-efficient solution for domain adaptation using finetuning.

\textbf{Finetuning Methodology.}
We fine-tuned the Phi-3.5-mini, Phi-3-mini, and Phi-3-medium models for the root cause analysis (RCA) task using Hugging Face's Supervised Fine-Tuning (SFT) Trainer, an efficient framework for adapting pretrained models to domain-specific datasets. From the incident dataset which we curated, we allocated 10K datapoints for validation, 2,891 for testing, and utilized the remainder for training (160K+). The data was split temporally, with older incidents designated for training and more recent cases for testing. This temporal split was implemented to simulate real-world scenarios, ensuring that the models could effectively generalize historical knowledge to newer, unseen incidents, reflecting the applicability to evolving system behaviors. 
The finetuning process was carried out over three epochs, utilizing a batch size of 64 samples on a compute cluster featuring 8 x NVIDIA Tesla V100 (672 GB RAM) and 8  x NVIDIA A100 (900 GB RAM) GPUs, depending on availability. To enhance model performance and generalization, we employed the AdamW optimizer, incorporating weight decay to mitigate the risk of over-fitting. Additionally, a linear learning rate scheduler with a warm-up phase was implemented to promote training stability and prevent abrupt changes in learning rates that could hinder convergence. The training duration for the models is given in \cref{tab:finetuning_time}.

\begin{table}[hbpt]
\caption{Training Duration: This table summarizes the total training times for small language models utilized in the Root Cause Analysis (RCA) task, highlighting model size and computational resources employed.}
\centering
\begin{adjustbox}{max width=\linewidth}
\begin{tabular}{@{}l c c c@{}}
    \toprule
    \textbf{Model} & \textbf{Model Size (Parameters)} & \textbf{Compute Type} & \textbf{Training Time (Hours)} \\ 
    \midrule
    Phi-3-mini & 3.8B & 8 x A100 & 6.5 \\
    Phi-3.5-mini & 3.8B & 8 x V100 & 13.5 \\
    Phi-3-medium & 14B & 8 x V100 & 30 \\
    \bottomrule
\end{tabular}
\end{adjustbox}
\label{tab:finetuning_time}
\end{table}

\textbf{Root Cause Generation.}
After finetuning the SLMs on the incident dataset, we performed inference and evaluated the models using a test set comprising of more recent incidents. This approach closely simulates real-world scenarios where models must leverage historical knowledge to analyze and address newer, unseen cases. By employing a temporal split—training on older incidents and testing on more recent ones—we assess the model's ability to generalize across time and adapt to evolving system behaviors, a crucial requirement for dynamic, real-time environments such as root cause analysis (RCA).

To ensure reproducibility and minimize randomness in the generated responses, we set the temperature to approximately zero during inference. A low temperature encourages the model to produce deterministic and focused outputs by reducing variation in token selection, which is essential when generating structured responses such as root cause explanations. Additionally, we capped the maximum number of new tokens at 200, aligning the response length with both the readability requirements of OCEs and the average token length of the ground truth root cause analyses. 

In the next section, we describe the performance of fine-tuned SLMs on the incident data and demonstrate how prompt instruction optimization enhances SLMs' ability to generate high-quality RCA recommendations. We detail the comparative results, showing the impact of optimized prompts on SLM accuracy and the quality of root cause identification, further validating the effectiveness of the optimization process.

\section{Experimental Setup}
In this section, we explain the default configurations used to tune the PromptWizard module, followed by explaining different versions of eARCO and baseline methods, and finally explain the evaluation strategy and performance metrics.   
\subsection{PromptWizard Configurations}
\label{sec:conf}
As discussed earlier, PromptWizard uses several configurable parameters to balance exploration and exploitation efficiently, ensuring that prompt optimization remains robust for RCA task. To derive an optimized prompt, we sample 25 random input-output pairs from the IcM dataset as training data. The following configuration was used for the optimization process:
\begin{itemize}
\item \textit{mutate\_refine\_iterations} : 3
    \item \textit{mutation\_rounds} : 3
    \item \textit{refine\_task\_eg\_iterations} : 3
    \item \textit{questions\_batch\_size} : 5
    \item \textit{min\_correct\_count} : 3
    \item \textit{few\_shot\_count} : 10
\end{itemize}
\subsection{Methods and Baselines}
\label{section:methods and baslines}
To thoroughly assess the impact of prompt optimization on both LLMs and finetuned SLMs, we experiment the following 8 strategies: 

    \par \textbf{Manual Prompt with Semantically Similar (SS) ICL examples \cite{zhang2024automated} - (Manual-SS):} The prompt, which is also proposed in \cite{zhang2024automated}, is designed with three key components: \textit{Default manual Instructions}, \textit{In-context Examples}, and \textit{Incident Details}. The \textit{Default manual Instructions}, includes the prompt hand-designed by a domain expert which has not undergone any optimization. This prompt briefly guides the model to assume the role of an OCE tasked with performing root-cause analysis (RCA) for Cloud incidents. For the \textit{In-context Examples}, we retrieve the top 10 similar incidents using the RAG pipeline, as detailed in \cref{subsection:ICL}, and provide their corresponding titles, summaries, owning service names, and ground truth root causes. Finally, we incorporate the details of the current incident, including its title, summary, and owning service name. This method is applied to both GPT-4, GPT4o and base SLMs. We deliberately exclude the finetuned models from this experiment, as including the In-context examples defeats the purpose of finetuning the models in the first place.

    \par \textbf{Optimized Prompt with Static examples- (PW-Default)}: In this configuration, we use \textit{PW Instructions} and \textit{PW static Examples} along with the \textit{Incident Details} in the prompt. This is an out-of-the box usage of PromptWizard without any modifications and the instructions and examples remain the same for all test incidents. This setup shows how well ICL examples selected by PW generalizes across all incidents.
    
    
    \par \textbf{Optimized Prompt with Semantic Similar examples (PW-SS)}: In this configuration, we incorporate only the \textit{PW Instructions}, excluding the \textit{PW static Examples}. The ICL examples are selected based on the semantic similarity of the incident at hand in run-time as described previously. This serves as an ablation study to isolate and assess the impact of the instructions themselves, allowing us to later compare and understand the contribution of the \textit{PW static Examples} to the overall performance.

    \par \textbf{Finetuned SLM (FtSLM):} This method utilizes the finetuned SLMs with a standard, non-optimized prompt that contains the \textit{Incident Title}, \textit{Incident Summary}, and \textit{Owning Service Name} in the same format used during the fine-tuning process. This serves as a benchmark for evaluating the performance of finetuned models without any further prompt optimization.
    
    \par \textbf{Finetuned SLM with PW Inst. \& Ex (FtSLM PW):} In this configuration, we augment the finetuned SLMs with both the \textit{PW Instructions} and \textit{PW static Examples}, along with the \textit{Incident Details} during inference. This allows us to assess whether the optimized instructions and static examples improve the performance of the finetuned models. 
    
    \par \textbf{Finetuned SLM with PW Inst.(FtSLM PW noEx.): } As another ablation experiment, we provide only the \textit{PW Instructions} along with the \textit{Incident Details}, without including the \textit{PW static Examples}. This allows us to evaluate the isolated impact of optimized instructions on the performance of the fine-tuned models.
    
    \par \textbf{Base SLM with PW Inst. \& Ex (BaseSLM PW): } In this baseline, we investigate the performance of the base (non-finetuned) SLMs, when equipped with both the \textit{PW Instructions} and \textit{PW static Examples}, excluding any finetuning. 
    
    \par \textbf{Base SLM with PW Inst (BaseSLM PW noEx.).: } This configuration uses the base SLMs with only the \textit{PW Instructions} and \textit{Incident Details}, omitting the \textit{PW static Examples}. This method seeks to compare the impact of finetuning and \textit{PW static Examples}.
    

\subsection{Evaluation Metrics}
Even with ground truth root cause information, evaluating the recommendations generated by language models is a complex task. While expert evaluations from the OCEs would provide the most accurate assessments, conducting such large-scale human evaluation is not a practical option due to time constraints of OCEs. Moreover, traditional automatic metrics, whether lexical or semantic, often fail to capture the nuanced, domain-specific similarities required for efficient evaluation as shown in previous studies \cite{zhang2024automated, goel2024x}. To address these challenges, we implement a dual evaluation strategy to assess the similarity and accuracy between the automatically generated recommendations and ground truth root causes: (1) an automated assessment using GPT-4 as the judge across the entire test dataset consisting of 2900 incidents; and (2) a small-scale human evaluation on a representative subset of incidents. 


\subsubsection{Automated Evaluation Using GPT-4} 

In this evaluation strategy, the GPT-4 model is prompted with a structured task description that specifies it's role as a scorer, tasked with comparing a generated string with a reference string based on a defined set of criteria. The model assigns a score between 1 to 5, where a higher score indicates a closer match in terms of content coverage, nuance, and accuracy. The model also provides a justification for each score. In addition to the reference and generated strings, we also provide the GPT-4 model with the incident summary as contextual information, allowing the model to evaluate the generated responses with a clearer understanding of the incident.


\subsubsection{Human Evaluation}

As incident owners are domain experts and have specific knowledge about the root cause context, we chose a subset of recent incidents, and interviewed the respective incident owners (on-call engineers). Along with the OCEs, we also asked other researchers (who are not incident management domain experts) to score two sets of incidents. Each evaluator was asked to score the generated responses based on two key criteria: (1) accuracy in comparison to the ground truth root cause and (2) readability in terms verbosity, grammatical correctness and structure of the recommendations. With detailed guidelines provided to ensure consistent scoring, each model-generated recommendation was rated on a scale of 1 to 5 for both criteria.



\section{Experimental Results}

\subsection{Evaluating Performance of eARCO with Large Models}

In this section, we compare the performance of manually crafted prompts with those optimized using PromptWizard for the RCA task. Table \ref{tab:performance-metrics-PW} presents the results across different configurations, with the LLM name indicating the model used for optimization and answer generation. The table reports the performance across two datasets: the `Complete Test Dataset' and the `Filtered Test Dataset'. The latter is a subset of the former, containing only those incidents where an incident summary is available in the \textit{Incident Details}. 

The optimized prompt generated by PromptWizard, denoted as \textit{PW-Default}, achieves 
average scores of 2.07 for GPT-4 and 2.13 for GPT-4o, outperforming the manually designed \textit{Manual-SS} prompts, which scored 2.03 and 2.07 for the same models, respectively.
Despite using the same 10 in-context examples for all test instances, PromptWizard's optimization outperforms \textit{Manual-SS}, which dynamically selects the 10 most semantically similar examples for each test instance. This result highlights the limitations of manually crafted prompts and the advantage of PromptWizard's automated optimization. Moreover, replacing the static in-context examples in PromptWizard with semantically similar examples \textit{(PW-SS)} further improves the performance to 2.33 and 2.51 for GPT-4 and GPT-4o, respectively, providing a massive 21\% gain in accuracy over manually designed prompts.

\begin{table}[hbpt]
\caption{GPT-4 evaluation scores for PromptWizard}
\centering
\begin{adjustbox}{max width=\linewidth}
\begin{tabular}{@{}l c c@{}}
    \toprule
    \textbf{Experiment}& \textbf{Complete Test Dataset}& \textbf{Filtered Test Dataset}\\ 
    \midrule
    \multicolumn{3}{c}{\textbf{GPT-4}} \\ 
    \midrule
    Manual-SS & 2.03 \scriptsize{$\pm$ 0.93} & 2.35 \scriptsize{$\pm$ 0.94}\\ 
    PW-Default   & 2.07 \scriptsize{$\pm$ 0.91} & 2.37 \scriptsize{$\pm$ 0.92} \\ 
    PW-SS & \textbf{2.33 \scriptsize{$\pm$ 0.98}} & \textbf{2.68 \scriptsize{$\pm$ 0.98}} \\
    \midrule
    \multicolumn{3}{c}{\textbf{GPT-4o}} \\ 
    \midrule
    Manual-SS &  2.07 \scriptsize{$\pm$ 1.01} & 2.33 \scriptsize{$\pm$ 1.05} \\ 
    PW-Default &  2.13 \scriptsize{$\pm$ 0.97} & 2.41 \scriptsize{$\pm$ 0.95} \\ 
    PW-SS & \textbf{2.51 \scriptsize{$\pm$ 1.01}} & \textbf{2.91 \scriptsize{$\pm$ 1.01}} \\
    \bottomrule
\end{tabular}
\end{adjustbox}
\label{tab:performance-metrics-PW}
\end{table}
To assess the value of PromptWizard's multi-step optimization process, we performed an ablation study where prompts were evaluated at intermediate stages of optimization (as shown in Table~\ref{tab:ablation-steps}). Results show a consistent improvement in performance as the prompt undergoes more stages of refinement, underscoring the importance of the multi-step optimization employed by PromptWizard. This confirms that each stage—mutation, scoring, critiquing, and synthesizing—contributes significantly to achieving optimal performance.

\begin{table}[hbpt]
\caption{GPT-4 evaluation scores for PW prompt from GPT-4o at various steps in the optimization process}
\centering
\begin{adjustbox}{max width=\linewidth}
\begin{tabular}{@{}l c c@{}}
    \toprule
    \textbf{Optimization Stage}& \textbf{Complete Test Dataset}& \textbf{Filtered Test Dataset}\\ 
    \midrule
    Base Prompt (Manual-SS)  &  2.07 \scriptsize{$\pm$ 1.01} & 2.33 \scriptsize{$\pm$ 1.05} \\
    After Prompt Instruction Tuning & 2.10  \scriptsize{$\pm$ 0.91} &  2.22 \scriptsize{$\pm$ 0.95}  \\ 
    After In-Context Examples Tuning  & 2.22   \scriptsize{$\pm$ 1.05} &  2.30 \scriptsize{$\pm$ 0.91}  \\ 
    PW Final Prompt (PW-SS) & \textbf{2.51 \scriptsize{$\pm$ 1.01}} & \textbf{2.91 \scriptsize{$\pm$ 1.01}}  \\ 
    \bottomrule
\end{tabular}
\end{adjustbox}
\label{tab:ablation-steps}
\end{table}


\begin{table}[h]
\caption{Results for base and fine-tuned SLMs}
\centering
\begin{adjustbox}{max width=\linewidth}
\begin{tabular}{@{}l c c@{}}
    \toprule
    \textbf{Experiment}& \textbf{Filtered Test Dataset}& \textbf{Complete Test Dataset}\\ 
    \midrule
    \multicolumn{3}{c}{\textbf{Phi-3.5-mini-128k-instruct}} \\ 
    \midrule
    FtSLM& 2.09 \scriptsize{$\pm$ 0.90} & 1.79 \scriptsize{$\pm$ 0.87} \\ 
    FtSLM PW noEx. & 2.13 \scriptsize{$\pm$ 0.87} & 1.90 \scriptsize{$\pm$ 0.87} \\ 
    FtSLM PW & \textbf{2.37 \scriptsize{$\pm$ 0.79}} & \textbf{2.01 \scriptsize{$\pm$ 0.84}} \\ 
    BaseSLM PW & 2.26 \scriptsize{$\pm$ 0.71} & 1.93 \scriptsize{$\pm$ 0.77} \\ 
    BaseSLM PW noEx. & 2.14 \scriptsize{$\pm$ 0.61} & 1.82 \scriptsize{$\pm$ 0.69} \\
    Manual-SS - BaseSLM & 1.79 \scriptsize{$\pm$ 0.58} & 1.55 \scriptsize{$\pm$ 0.60}\\
    \midrule
    \multicolumn{3}{c}{\textbf{Phi-3-medium-128k-instruct}} \\ 
    \midrule
    FtSLM& 2.11 \scriptsize{$\pm$ 0.99} & 1.82 \scriptsize{$\pm$ 0.93} \\ 
    FtSLM PW noEx. & 2.17 \scriptsize{$\pm$ 0.84} & 1.87 \scriptsize{$\pm$ 0.86} \\ 
    FtSLM PW & \textbf{2.21 \scriptsize{$\pm$ 0.86}} & \textbf{1.93 \scriptsize{$\pm$ 0.90}} \\ 
    BaseSLM PW & 2.00 \scriptsize{$\pm$ 0.44} & 1.68 \scriptsize{$\pm$ 0.57} \\ 
    BaseSLM PW noEx. & 2.01 \scriptsize{$\pm$ 0.58} & 1.69 \scriptsize{$\pm$ 0.65} \\
    Manual-SS - BaseSLM & - & - \\
    \midrule
    \multicolumn{3}{c}{\textbf{Phi-3-mini-128k-instruct}} \\ 
    \midrule
    FtSLM& 2.08 \scriptsize{$\pm$ 0.99} & 1.79 \scriptsize{$\pm$ 0.92} \\ 
    FtSLM PW noEx. & 1.98 \scriptsize{$\pm$ 0.74} & 1.77 \scriptsize{$\pm$ 0.74} \\ 
    FtSLM PW & \textbf{2.12 \scriptsize{$\pm$ 0.84}} & \textbf{1.83 \scriptsize{$\pm$ 0.84}} \\ 
    BaseSLM PW & 1.66 \scriptsize{$\pm$ 0.59} & 1.44 \scriptsize{$\pm$ 0.67}\\ 
    BaseSLM PW noEx. & 2.10 \scriptsize{$\pm$ 0.63} & 1.74 \scriptsize{$\pm$ 0.70} \\
    Manual-SS - BaseSLM & 1.83 \scriptsize{$\pm$ 0.66} & 1.58 \scriptsize{$\pm$ 0.66} \\

    \bottomrule
\end{tabular}
\end{adjustbox}
\label{tab:performance-metrics}
\end{table}

\subsection{Evaluating Performance of eARCO with Small Models}

As discussed earlier, SLMs demonstrate significant potential when finetuned for domain specific tasks such as RCA. By leveraging such models, organizations can reduce the overhead costs associated with querying expensive LLMs or maintaining large retrieval corpora for RAG pipelines.
In this section, we present the evaluation results of the responses generated by these SLMs. Furthermore, we demonstrate how the prompt-optimization framework improves the performance of these models, enhancing both accuracy and efficiency.

The results presented in Table \ref{tab:performance-metrics} provide a detailed comparison of various finetuning and prompting strategies for SLMs, as outlined in \cref{section:methods and baslines}. Specifically, the three models—Phi-3-medium, Phi-3-mini, and Phi-3.5-mini—were evaluated under different configurations to examine the impact of \textit{finetuning}, \textit{PW Instructions}, and \textit{PW static Examples} on model performance. 

The table presents performance across two datasets: the \textit{Complete Test Dataset} and the \textit{Filtered Test Dataset}, where the latter only includes incidents with available summaries in the \textit{Incident Details}. Incident summaries provide crucial context for accurate RCA generation, and their absence can hinder model performance. When evaluating the Filtered Test Dataset, we observe a consistent improvement in accuracy and relevance. This highlights the critical role of contextual information in enhancing RCA task performance for SLMs.



Across all three models (Phi-3.5-mini, Phi-3-medium, Phi-3-mini), the finetuned models with \textit{PW Instructions} and \textit{PW static Examples} consistently achieve the highest scores on both the filtered and complete test datasets. For instance, the Phi-3.5-mini model shows the highest performance amongst all the models, with an average score of 2.37 on the filtered test dataset and 2.01 (\textit{FtSLM PW}) on the complete test dataset, indicating that optimized prompts significantly enhance the model's ability to predict accurate root causes.

However, when comparing the performance of finetuned SLMs without the \textit{PW static Examples} (\textit{FtSLM PW noEx.}), there is a noticeable decrease in scores. For instance, in the Phi-3.5-mini model the scores drop to 2.13 on the filtered dataset, and to 1.90 on the complete dataset. This demonstrates the importance of providing the static examples (\textit{PW static Examples}) to the finetuned SLMs to impart crucial reasoning and domain-specific knowledge, without incurring additional dynamic example retrieval cost.

The base SLM models with PromptWizard instructions also perform reasonably well, though they lag behind the finetuned models. For instance, the Phi-3.5-mini \textit{BaseSLM PW} achieves 2.26 on the filtered dataset, showing improvement over the base model under the \textit{Manual-SS} setting (which scored 1.79). Nevertheless, the performance gap between the base and finetuned SLM is evident, especially when prompt optimization with examples is applied. 

Lastly, applying the \textit{Manual-SS} configurations to the base SLMs demonstrate the lowest performance across all settings. These models, without finetuning or prompt engineering, score the lowest on both datasets, reflecting the baseline performance before any optimizations are applied. In particular, the Phi-3-medium model under this configuration produces in null responses, underscoring the limited effectiveness of the base models without further enhancements.

\subsection{Ablation Results}
To analyse the utility of the semantically similar in-context examples, we perform an ablation by varying the number of examples in the prompt with PromptWizard instructions. Table \ref{tab:ablation-icl} shows performance of the prompt with number of example ranging from 0 to 10. We observe that increasing the number of in-context examples leads to better performance on the evaluation set going from 1.97 for the zero-shot setting to 2.51 for the 10-shot setting on the complete dataset and from 2.13 to 2.91 on the filtered dataset. Hence we observe around 27\% improvement on the complete evaluation set and 37\% improvement on the filtered evaluation set dataset.

\begin{table}[hbpt]
\caption{GPT-4 evaluation scores for ablation on number of semantically similar In-Context Examples with PW Instructions}
\centering
\begin{adjustbox}{max width=\linewidth}
\begin{tabular}{@{}l c c@{}}
    \toprule
    \textbf{Number of Examples}& \textbf{Complete Test Dataset}& \textbf{Filtered Test Dataset}\\ 
    \midrule
    \multicolumn{3}{c}{\textbf{GPT-4o}} \\ 
    \midrule
    0 &  1.97 \scriptsize{$\pm$ 0.91} & 2.13 \scriptsize{$\pm$ 0.88} \\ 
    3 &  2.07 \scriptsize{$\pm$ 0.98} & 2.25 \scriptsize{$\pm$ 0.90} \\ 
    5 &  2.24 \scriptsize{$\pm$ 1.02} & 2.41 \scriptsize{$\pm$ 0.91} \\ 
    7 &  2.40 \scriptsize{$\pm$ 0.97} & 2.72 \scriptsize{$\pm$ 0.94} \\ 
    10 & \textbf{2.51 \scriptsize{$\pm$ 1.01}} & \textbf{2.91 \scriptsize{$\pm$ 1.01}} \\
    \bottomrule
\end{tabular}
\end{adjustbox}
\label{tab:ablation-icl}
\end{table}

\subsection{Human Evaluation Results}
To understand the concrete impact of eARCO, we reached out to 47 OCEs involved in recent incident resolutions to assess responses generated by various models. These models included GPT-4 and GPT-4o under the \textit{Manual-SS} settings, GPT-4o using the \textit{PW-SS} configuration, and the Phi-3.5-mini model with the \textit{FtSLM PW} setup. Table \ref{tab:evaluation_metrics} illustrate the accuracy and readability scores assigned by the OCEs. 

The GPT-4o model with optimized PromptWizard instructions and 10 semantically similar examples (PW-SS) achieved the highest average accuracy score of 2.91, reflecting a 14.12\% and 7.45\% improvement over GPT-4 and GPT-4o using Manual-SS, respectively. 
For readability, the GPT-4o model in both Manual-SS and PW-SS configurations scored highly, with a rating of 4.21, showing a 3.19\% improvement over GPT-4 under Manual-SS. Lastly, despite being scalable and cost-effective solution, finetuned SLMs with optimized instructions (FtSLM-PW) achieves an average accuracy of 2.23 and even provided better recommendations than LLMs for a small number of incidents, which suggests that FtSLM-PW can be an attractive alternative cost-effective option for RCA generation task.

In addition to evaluations from OCEs, we gathered scores from 10 researchers who are not incident management domain experts, but have access to the ground truth root causes and model recommendations. These human evaluators are divided into two groups. Each group evaluated 25 incidents, resulting in a total of 50 incidents. Table \ref{tab:RF_eval} presents the aggregated accuracy and readability scores. Consistent with the OCEs’ evaluation, PW-SS achieves the highest accuracy score of 3.50 and the highest readability score of 4.30.

\begin{table}[hbpt]
    \centering
    \caption{Accuracy and Readability scores assigned by OCEs}
    \begin{adjustbox}{max width=\linewidth}
    \begin{tabular}{@{}l c c@{}}
        \toprule
        \textbf{Model} & \textbf{Accuracy} & \textbf{Readability} \\
        \midrule
        GPT-4 & 2.55  \scriptsize{$\pm$1.26} & 4.08  \scriptsize{$\pm$0.90} \\
        \midrule
        GPT-4o & 2.74  \scriptsize{$\pm$1.42} & \textbf{4.21  \scriptsize{$\pm$0.90}} \\
        \midrule
        PW-SS & \textbf{2.91  \scriptsize{$\pm$1.36}} & \textbf{4.21  \scriptsize{$\pm$0.95}} \\
        \midrule
        FtSLM PW & 2.23  \scriptsize{$\pm$1.30} & 3.68  \scriptsize{$\pm$1.23} \\
        \bottomrule
    \end{tabular}
    \end{adjustbox}
    \label{tab:evaluation_metrics}
\end{table}

\begin{table}[hbpt]
    \centering
    \caption{Accuracy and Readability scores assigned by domain experts}
    \begin{adjustbox}{max width=\linewidth}
    \begin{tabular}{@{}l c c@{}}
        \toprule
        \textbf{Model} & \textbf{Accuracy} & \textbf{Readability} \\
        \midrule
        GPT-4 & 3.15  \scriptsize{$\pm$1.25} & 3.93  \scriptsize{$\pm$0.79} \\
        \midrule
        GPT-4o & 3.38  \scriptsize{$\pm$1.20} & 4.06  \scriptsize{$\pm$0.82} \\
        \midrule
        PW-SS & \textbf{3.50  \scriptsize{$\pm$1.20}} & \textbf{4.30  \scriptsize{$\pm$0.64}} \\
        \midrule
        FtSLM PW & 2.89  \scriptsize{$\pm$1.13} & 3.86  \scriptsize{$\pm$0.65} \\
        \bottomrule
    \end{tabular}
    \end{adjustbox}
    \label{tab:RF_eval}
\end{table}



\section{Discussion}

\textbf{Promise of prompt optimization.} Prompt instruction optimization has gained popularity recently and proven to be effective in many traditional NLP tasks. In this work, we first analyse and demonstrate that optimized instructions can significantly improve the performance on sensitive proprietary domain such as incident management as well. Furthermore, we demonstrate that finetuned domain adapted SLMs can be an attractive alternative for AIOPs tasks. Moreover, we demonstrate that the optimized instructions and static diverse synthetic in-context examples can significantly boost the performance of domain adapted fintuned SLMs. These insights will be valuable for designing more efficient and cost-effective AIOPs solutions in future.

\textbf{Deployment status and scale.} We conducted our experiments and evaluation by leveraging data from IcM of \CompanyX{}. We have deployed a large-scale RCA recommendation system as a service by leveraging ICL pipeline with LLMs that is serving more than hundred internal service teams for more than six months. For our experiments, the recommendations for Manual-SS with GPT-4 method are directly obtained from production environment. We have also recently deployed the finetuned SLM solution to augment the recommendations generated by the ICL pipeline. As eARCO demonstrate superior performance on real-world production incident dataset, we plan to deploy eARCO for both the ICL and finetuned SLM in near future.    

\textbf{Threats to validity.} Although the optimized prompt instruction demonstrate superior performance with different settings of LLMs and finetuned SLMs, the performance of eARCO is highly dependant on the underlying language model. As we have evaluated the performance of our models on incident data from \CompanyX{} only, the performance may vary if evaluated on different dataset from other organizations. While the GPT-4 based accuracy evaluation has been widely adopted, these evaluation results can be slightly noisy due to hallucination problem of LLMs. Moreover, the human evaluation is conducted on a small set of incidents as it is challenging to and time consuming to scale these feedback collection process. In some cases, the model may potentially generate hallucinated responses and additional noisy mitigation suggestions, which can be problematic and misguide incident owners. Lastly, we have only finetuned Phi-series of SLMs, but the performance might improve further with other open-source SLMs. Moreover, we plan to use domain adaptation techniques using RLHF techniques to improve the performance of SLMs in future.



\section{Related Work}
In this section, we summarize the existing work on incident management, adoption of LLMs for RCA and AIOPs tasks, and existing prompt optimization techniques. 
\subsection{Incident Management}







Given its practical importance, AI for Operations (AIOPs) techniques have become popular for automatically resolving issues arising from different stages of incident lifecycle management. Empirical studies have been adopted broadly to understand the gaps and limitations in existing large-scale cloud services, either delving into the types incident root causes~\cite{gao2018empirical,zhang2021understanding} or system-level issues~\cite{liu2019bugs,ghosh2022fight}. Several AIOPs techniques have been proposed to address the challenges in detection \cite{ganatra2023detection, Srinivas2024IntelligentMonitoring}, triaging \cite{EmpiricalIcMICSE2019}, diagnosis \cite{bansal2019decaf}, and mitigation \cite{jiang2020mitigate} to either reduce human efforts or accelerating the incident resolution process. Our work enhances the performance of automated root cause generation task where accuracy and efficiency are primary goals.

\subsection{Promise of LLMs in Incident Management}
In large-scale cloud services, effective and efficient handling of incidents is essential. Given the superior performance of LLMs in several domain-specific software engineering tasks ranging including code generation \cite{chen2021evaluating, xu2022systematic}, program synthesis(\cite{jain2022jigsaw}), code review \cite{li2022auger, li2022automating}, code repair \cite{wadhwa2024core,joshi2023repair} and code-fix \cite{jimenez2023swe}, LLMs have been adopted increasingly for solving problems in incident management domain. 
Several recent works propose to address the incident diagnosis and RCA \cite{ahmed2023recommending, chen2024automatic, zhang2024automated} tasks using LLMs. Ahmed \textit{et. al.}~\cite{ahmed2023recommending} propose to finetune a GPT model for learning domain specific knowledge about incident management and recommend potential root causes at the time of incident creation. Zhang \textit{et. al.}~\cite{zhang2024automated} subsequently propose an efficient RAG based in-context learning method for RCA generation task. In contrast to the existing works that leverages manually designed static instructions with LLMs, we propose to identify an optimized prompt instructions with synthetic in-context examples and further demonstrate the potential of cost-effective finetuned SLMs for RCA generation tasks.


\subsection{Prompt Optimization}
Existing LLM-based solutions for incident management heavily rely on prompt engineering techniques. The prompts are carefully chosen, often with multiple trials and errors, based on their effectiveness on the task being solved. This requires manual effort and generated prompts could be sub-optimal due to lack of systematic exploration of prompts. Furthermore, the performance of static prompt varies significantly as the underlying LLM evolves. Prompt optimization techniques address these limitations by optimizing either soft-prompt \cite{chen2023instructzero, linuse} or candidate prompts \cite{fernando2023promptbreeder, zhou2022large, guo2023connecting}. PromptWizard \cite{agarwal2024promptwizard} is a recent work that outlines and addresses the limitations of existing approaches - (i) high computation cost, (ii) lack of human-interpretable prompts, (iii) sub-optimal prompt outputs for complex tasks, (iv) lack of feedback-based exploration. Given its efficiency and superior performance on several traditional NLP tasks, we leverage PromptWizard in this work for identifying optimized prompt instruction and synthetic in-context examples.

\section{Conclusion}
In this work, we propose eARCO, an efficient and scalable optimized framework for automatically generating accurate root cause recommendations for incidents in large-scale cloud services. 
By leveraging state-of-the-art prompt optimization technique, we automatically identified optimized prompt instruction that is augmented with dynamically retrieved semantically similar in-context examples during real-time inference. Moreover, we develop scalable and cost-effective finetuned SLMs and demonstrate that, with optimized prompt instructions, these can be an attractive alternative solution for RCA generation task. 
Our extensive experimental evaluation by domain experts and GPT-4 as judge demonstrate that eARCO improves the accuracy of RCA recommendations significantly for both LLMs and finetuned SLMs on real-world incidents from \CompanyX. We believe these insights will motivate the need of prompt optimization and adoption of cost-effective finetuned SLMs in solving various challenges arising from different stages of incident management lifecycle.

\balance
\bibliographystyle{ACM-Reference-Format}
\bibliography{Reference}

\end{document}